\documentclass[letterpaper,11pt,reqno]{amsart}

\newcommand{\sets}[1]{\left\{{#1}\right\}} 

\theoremstyle{plain}

\theoremstyle{definition}

\theoremstyle{remark}

\usepackage{enumerate}
\usepackage{amssymb}
\usepackage[mathscr]{eucal}
\usepackage{graphicx,color}

\usepackage{hyperref}
\usepackage{interval}
\usepackage{float}
\usepackage{multicol}
\usepackage[table,xcdraw]{xcolor}
\usepackage{multirow}
\usepackage{graphicx}
\usepackage{subcaption}

\intervalconfig{soft open fences}

\pdfoutput=1

\begin{document}

\title{Real-time on-device nod and shake recognition}

\author[Elmar H. Langholz, Reuben Brasher]{
Elmar H. Langholz, Reuben Brasher\\
\MakeLowercase{\{\href{mailto:elan@microsoft.com}{elan}, \href{mailto:rebrashe@microsoft.com}{rebrashe}\}@microsoft.com
}}

\date{\today}
\maketitle

\begin{abstract}
We discuss methods for teaching systems to identify gestures such as head nod and shake. We use iPhone X depth camera to gather data and later use similar data as input for a working app. These methods have proved robust for training with limited datasets and thus we make the argument that similar methods could be adapted to learn other human to human non-verbal gestures. We showcase how to augment Euler angle gesture sequences to train models with a relatively large number of parameters such as LSTM and GRU and gain better performance than reported for smaller models such as HMM. In the examples here, we demonstrate how to train such models with Keras and run the resulting models real time on device with CoreML.
\end{abstract}

\section{Introduction}

As new technologies come in, we as users have to learn, adapt and understand how to use these new technologies as tools. However, it need not be only we who adapt to the tools, but the tools could also adapt to us with the application of machine learning. There has been significant work on applying machine learning to gesture recognition in general, but much of the focus of such work has been on interfaces implemented with touch, typing and speech. This leaves opportunities to explore interactions using non-verbal gestures detected by portable device cameras. Such interactions are particularly attractive now that mobile devices are almost ubiquitous and for many people a primary tool for task completion and communication. Hence in this paper we will explore techniques to teach machines to identify previously unseen gestures from a few sample interactions similar to interactions already prevalent in human to human communication.

In \cite{kawato2000real}, Kawato and Ohya first introduced a method of detecting a head nod and shake through a webcam video stream by tracking a mid-point in the head between the eyes through circle-frequency filtering with skin color and templating. By using infrared pupil tracking and discrete Hidden Markov Model (HMM), Kapoor and Picard obtained an accuracy of $78.46\%$ \cite{kapoor2001real}. Similarly, Davis and Vaks \cite{davis2001perceptual} used IBM's PupilCam to obtain face location and used a timed finite state machine to detect nod and shake. Tan and Rong \cite{tan2003real} obtained accuracy of $85\%$ by using AdaBoost face detector and eye location jointly with a discrete HMM. The authors of \cite{kwon2006cylindrical} exposed a new method in which they used a 3D cylindrical head model with dynamic template and accumulative HMM's. Wei et al. \cite{wei2013real} used a Microsoft's Kinect depth sensor to obtain an accuracy of $86\%$ by using the head pose angles with three HMM's. Finally, the authors of \cite{chen2015head} proposed a new method for head nod detection in which they leveraged a full 3D face centered rotation model demonstrating its validity in two and four party conversations.

In this paper we will focus on two types of head gestures: nodding and shaking. We demonstrate that a mobile device can understand a user approval and disapproval. To our knowledge this is the first time such an approach has taken advantage Apple's iPhone-X\cite{iphonex} depth-sensing front-facing camera, ARKit\cite{arkit} face tracking and CoreML\cite{coreml}.

\section{Data}

\subsection{Collection}

To collect gesture sequences from users, we built an iOS application which randomly prompted a user for three different types of gesture: nod, shake and other. For each prompt the app alloted time randomly between two to four seconds to record a gesture. The app showed a progress bar, with the intent of forcing a user to provide different speeds of a gesture. Before asking a user to provide sample gestures through the application, we primed them on the meaning of each gesture by providing examples. We also explained how to start and stop data collection but left the number of samples collected to their discretion. Upon stopping collection the app aggregated, compressed and uploaded data to a backend.

ARKit provides estimates head pose as Euler angles $\phi = \sets{x,\, y,\, z}$, where $x$ corresponds to pitch, $y$ to yaw and $z$ to roll. While ARKit tracked a user's face it provided samples at a frequency of $f = 60\, \text{Hz}$. However, when no face was detected no values were collected. This resulted in sequences of length between $0$ and $240$ samples. We collected $536$ sequences from $6$ users.

\begin{table}[H]
    \centering
    \caption{Gesture frequency}
    \label{table-gesture-frequency}
    \begin{tabular}{|c|c|c|}
        \hline
        \rowcolor[HTML]{C0C0C0} 
        \textbf{Nod} & \textbf{Shake} & \textbf{Other} \\ \hline
        187          & 163            & 186            \\ \hline
    \end{tabular}
\end{table}

\subsection{Analysis}

After performing data analysis on the collected data set, the other state turned out to be a lot more varied regarding its sequence shapes and values than nod and shake. We speculate that this is because our contributors understood more uniformly what nod and shake gestures should be in comparison to the more ambiguous other gesture. It turned out that many of the samples corresponding to the other gesture were almost motionless. In other words, data contributors usually decided just to hold the device and look at the screen. For this reason, instead of using the full set of Euler angles $\phi$, we decided that using $\phi_x$ (pitch) and $\phi_y$ (roll) was enough to be able to differentiate between the three types of gestures.

\subsection{Process}

We first filtered data to remove sequences of length less than $50$. Thus all remaining sequences had length between $50$ and $240$, which in time length correspond between $5/6$ and $4$ seconds. The collected data was split as follows:

\begin{table}[H]
    \centering
    \caption{Data split}
    \label{table-data-split}
    \begin{tabular}{|cccc|ccc|}
        \cline{2-7}
        \multicolumn{1}{l}{}                                        & \multicolumn{3}{c|}{\cellcolor[HTML]{C0C0C0}\textbf{Train}} & \multicolumn{3}{c|}{\cellcolor[HTML]{C0C0C0}\textbf{Test}} \\ \cline{2-7} 
        \multicolumn{1}{c|}{\cellcolor[HTML]{C0C0C0}\textbf{Size}}  & \multicolumn{3}{c|}{90\%}                                   & \multicolumn{3}{c|}{10\%}                                   \\ \hline
        \rowcolor[HTML]{EFEFEF} 
        \multicolumn{1}{c|}{\cellcolor[HTML]{C0C0C0}\textbf{Type}}  & \textit{Nod}      & \textit{Shake}     & \textit{Other}     & \textit{Nod}     & \textit{Shake}     & \textit{Other}     \\ \hline
        \multicolumn{1}{c|}{\cellcolor[HTML]{C0C0C0}\textbf{Count}} & 172               & 147                & 163                & 15               & 16                 & 23                 \\ \hline
        \multicolumn{1}{c|}{\cellcolor[HTML]{C0C0C0}\textbf{Total}} & \multicolumn{3}{c|}{482}                                    & \multicolumn{3}{c|}{54}                                    \\ \hline
    \end{tabular}
\end{table}

After splitting the data, we removed $\phi_z$ (yaw) for all samples as per our observation during data analysis.

\subsection{Augmentation}

We performed two types of data augmentation. The first type of augmentation was, general shrinking and stretching on a full sequence. The second type was shrinking and stretching heads and tails of a sequence. Shrinking and stretching amounts to downsampling and upsampling of a sequence range. We were aware that time warping and resampling is regularly frowned upon since it alters sequence speed perception especially in examples like activity from accelerometer data\cite{forsyth2018applied}. Nonetheless, this is exactly what we were trying to accomplish because we believe that human beings have more than a single speed for a gesture depending on different physical and emotional state. To be explicit, we wanted to demonstrate a methodology for learning gestures robust to having limited amounts of data.

Throughout the following, let $l(s)$ denote the length in samples of a sequence, $s$. For two sequences $s_1$ and $s_2$ denote the concatenation of the two sequences by
$$s_1 \oplus s_2$$

The operation of shrinking or stretching a full sequence consisted of resampling a sequence to a defined set of sizes. These sizes were calculated by using a step size of $\Delta_{\alpha} = 30$. For shrinking, we subtracted sequence length by positive integer multiple of $\Delta_{\alpha}$ as long as that the resulting length was positive. Similarly for stretching, we added integer multiples of $\Delta_{\alpha}$ as long as the resulting length was at most $240$.

For the second type of operation, shrinking and stretching heads and tails of sequences, first we actually had to determine what the heads and tails of the sequences were. In the case of nod and shake gestures, we used the Pelt algorithm\cite{killick2012optimal} to detect the signals initial $c_i$ and final $c_f$ change point. With these change points the head subsequence interval was $Q_{ns}^{h} = \interval[open right]{0}{c_i}$ and the tails subsequence interval was $Q_{ns}^{t} = \interval[open right]{c_f}{l(s)}$. For the other gesture, we simply considered the leading and trailing subsequences to be the first and last fifth of the sequence.

Let $R^{h}$ and $R^{t}$ be the remaining interval of a sequence. Thus, augmented sequences were one of the following:

$$Q^{h} \oplus R^{h}$$
$$R^{t} \oplus Q^{t}$$

Prior to concatenation we performed downsampling or upsampling on the values of sequence $s$ in interval $Q$. We augmented sequences in step sizes of $\Delta_{\beta} = -4$ for downampling and $\Delta_{\beta} = 4$ for upsampling. We stopped downsampling, once a $Q$ would have reached length zero or less. Similarly, we ended upsampling once $Q$  would have been length $240$ or more.

These two types of data augmentation led to having the following data set size breakdown:

\begin{table}[H]
    \centering
    \caption{Data counts after augmentation}
    \label{table-data-augmentation}
    \begin{tabular}{|cccc|ccc|}
        \cline{2-7}
        \multicolumn{1}{l}{}                                        & \multicolumn{3}{c|}{\cellcolor[HTML]{C0C0C0}\textbf{Train}} & \multicolumn{3}{c|}{\cellcolor[HTML]{C0C0C0}\textbf{Test}} \\ \cline{2-7} 
        \rowcolor[HTML]{EFEFEF} 
        \multicolumn{1}{c|}{\cellcolor[HTML]{C0C0C0}\textbf{Type}}  & \textit{Nod}      & \textit{Shake}     & \textit{Other}     & \textit{Nod}     & \textit{Shake}     & \textit{Other}     \\ \hline
        \multicolumn{1}{c|}{\cellcolor[HTML]{C0C0C0}\textbf{Count}} & 62356             & 55581              & 54757              & 5520             & 5893               & 7562               \\ \hline
        \multicolumn{1}{c|}{\cellcolor[HTML]{C0C0C0}\textbf{Total}} & \multicolumn{3}{c|}{172694}                                 & \multicolumn{3}{c|}{18975}                                 \\ \hline
    \end{tabular}
\end{table}

\subsection{Standardization}

We calculated the mean and variance of the training data, and normalized the data to zero mean and unit variance on both training and test data. Since the samples were of a variable length, we also padded all sequences with zeros at the end to length $240$.

\subsection{Flattening}

Finally, with the intent of being able to have a single contiguous sequence representing the bi-dimensional data instead of two disjoint sequences, we merged the two pitch $\phi_x$ and roll $\phi_y$ sequences corresponding to a sequence $s$ into a flattened sequence $s^f$ by performing the following operation:

$$s_{i}^{f} = s_{i - 1}^{f} \oplus (\phi_{x_i} \oplus \phi_{y_i})$$

In other words, we concatenated the individual values of pitch and roll corresponding to the same position and then concatenated these merged values.

\section{Modeling}

With enough data, we built classifiers using a recurrent neural networks. For this we used Keras\cite{chollet2015keras} and a highly customizable model interface allowing us to programmatically change network size and other hyper-parameters. As baseline an architecture consisting of a single memory layer followed by a dense layer and an activation layer. Effectively, the form of the constructed RNN was as follows:

\begin{figure}[H]
    \caption{RNN architecture}
    \centering
    \includegraphics[width=0.25\textwidth]{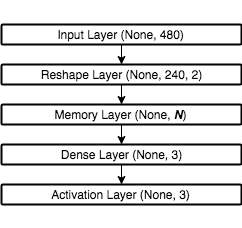}
\end{figure}

\subsection{Reshaping layer}

Since the data was normalized, standardized and flattened into a sequence we needed to be able to decouple it back into what is known in the computer vision community as channels. In our case we have two channels: pitch and yaw. The reshaping layer converts as input sequence size of $480$ into $(240, 2)$.

\subsection{Memory layer}

Consists of a defined number of connected memory cells (hidden units). There are different types of memory cells but the two popular ones that we used were long short-term memory (LSTM)\cite{hochreiter1997lstm} and gated recurrent unit (GRU)\cite{choalemnlp14}. For either memory cell type we used a tanh activation function and a hard sigmoid recurrent activation function. The memory layer converts input of shape $(240, 2)$ into a vector of length $N$, the number of hidden units.

\subsection{Dense and activation layer}

A densely-connected neural network layer of size $3$, one for each class: nod, shake and other. The dense layer followed by an activation layer using a softmax activation which maintains a size of $3$.

\section{Hyper-parameter search}

Having established a common architecture, we performed a hyper-parameter search to find optimal values for the memory layer. Specifically, a grid search using a 5-fold cross-validation of $10$ epochs each with a batch size of $80$ on the training data. For the search, hyper-parameters Table \ref{hyper-parameters}.

\begin{table}[H]
    \centering
    \caption{Hyper-parameters}
    \label{hyper-parameters}
    \begin{tabular}{|c|c|}
        \hline
        \rowcolor[HTML]{C0C0C0} 
        \textbf{Name} & \textbf{Values}         \\ \hline
        Memory cell   & LSTM, GRU               \\ \hline
        Hidden units  & 64, 128, 256, 512, 1024 \\ \hline
    \end{tabular}
\end{table}

In total, we trained ten models using an RMSProp optimizer with a sparse categorical cross entropy loss. The best accuracy was given by GRU memory cells with a layer consisting of $256$ hidden units.

\section{Results}

\begin{figure}[H]
    \centering
    \caption{Loss and accuracy}
    \label{training-loss-accuracy}
    \begin{subfigure}{0.45\textwidth}
        \centering
        \includegraphics[width=0.9\linewidth]{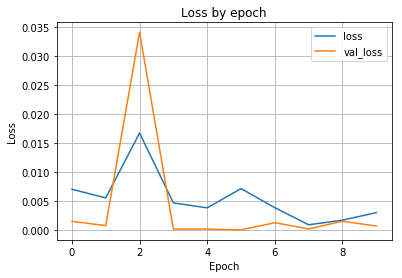}
        \caption{Training (blue) and validation (orange) loss for ten epochs}
    \end{subfigure}
    \begin{subfigure}{0.45\textwidth}
        \centering
        \includegraphics[width=0.9\linewidth]{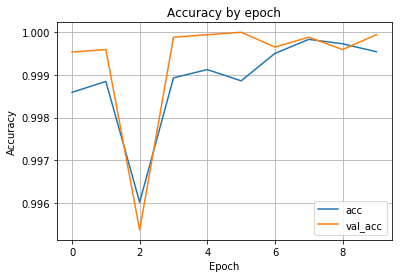}
        \caption{Training (blue) and validation (orange) accuracy for ten epochs}
    \end{subfigure}
\end{figure}

Having obtained the best hyper-parameters for the memory layer, we used these with the same configuration described above. However, this time we with-held $10\%$ of the training set as the validation set and retrained the model. The Figure \ref{training-loss-accuracy} show the training loss and accuracy per each epoch for the optimal model.

Evaluated on the test set the optimal model gave an accuracy of $91.78\%$. In order to provide further insight we provide the confusion matrix corresponding to each class.

\begin{table}[H]
    \centering
    \caption{Confusion matrix}
    \label{confusion-matrix}
    \begin{tabular}{cc|c|c|c|}
        \cline{3-5}
                                                                                  & \multicolumn{1}{l|}{}                  & \multicolumn{3}{c|}{\cellcolor[HTML]{EFEFEF}Predicted}                                                                                                                                   \\ \cline{3-5} 
                                                                                  & \multicolumn{1}{l|}{}                  & \multicolumn{1}{l|}{\cellcolor[HTML]{C0C0C0}\textbf{Nod}} & \multicolumn{1}{l|}{\cellcolor[HTML]{C0C0C0}\textbf{Shake}} & \multicolumn{1}{l|}{\cellcolor[HTML]{C0C0C0}\textbf{Other}} \\ \hline
        \multicolumn{1}{|c|}{\cellcolor[HTML]{EFEFEF}}                            & \cellcolor[HTML]{C0C0C0}\textbf{Nod}   & 4908                                                      & 587                                                         & 25                                                          \\ \cline{2-5} 
        \multicolumn{1}{|c|}{\cellcolor[HTML]{EFEFEF}}                            & \cellcolor[HTML]{C0C0C0}\textbf{Shake} & 0                                                         & 5618                                                        & 275                                                         \\ \cline{2-5} 
        \multicolumn{1}{|c|}{\multirow{-3}{*}{\cellcolor[HTML]{EFEFEF}Actual}} & \cellcolor[HTML]{C0C0C0}\textbf{Other} & 0                                                         & 672                                                         & 6890                                                        \\ \hline
    \end{tabular}
\end{table}

\subsection{On-device prediction}

We exported the Keras model into a CoreML model of $803\,\text{kb}$ in size. The resulting model integrated easily with the data collection application by adding a switch in order to swap between collection and prediction while keeping the same start/stop visual semantics. Reusing the same collection mechanism, we collected the real-time Euler Angles until we filled a buffer of $240$ samples or $4\,\text{s}$ of data. On filling the buffer, the app generated its first prediction and afterwards every $15$ time steps (or $\frac{1}{4}\,\text{s}$). In order to perform the prediction the buffer was standardized (using the trained model's standardization parameters), flattened and then fed into the RNN-GRU.

We provided the application to several of users.  While they were impressed, some did give the feedback that under certain circumstances, it felt sluggish. We attribute this to two factors. The first factor is that the initial $4\,\text{s}$ wait time until the buffer fills causes the user to feel that nothing is happening. This is easily solvable by padding with zeros at the front of the unfilled buffer and evaluating as if it were full.

The second factor is that because the buffer is large there are times that two or more gestures will be in the buffer at the same time. Under these circumstances a second gesture will often not be recognized until the first gesture has shifted out of the buffer. One solution is that because we trained our model to recognize one gesture at a time we could partially shift an already recognized gesture out of the buffer before starting to detect the next gesture. An alternative solution would be to make the buffer smaller and pad before evaluating the model. A third solution would be to further augment our data set by stitching observations together and shifting the buffer such that we use majority voting to assign a label.

\section{Future work}

In this paper, we have been able to show that performing nod and shake recognition on-device today is not only possible but easily done. The RNN-GRU model performs better than other previously reported HMM models but it requires significantly more data which is why we propose using data augmentation. In future work we would like to experiment with making on-device recognition faster, building RNN models with less data and personalizing nod and shakes such that we are able to detect different sub-types for a user. 

\bibliographystyle{amsplain}
\bibliography{reading}

\end{document}